\documentclass[12pt]{article}
\usepackage{a4}
\usepackage[dvips]{graphicx}
\usepackage{cite}
\usepackage{here}
\usepackage{rotating}
\newcommand{\as}   {\ensuremath{\alpha_{\rm S}}}
\newcommand{\yc}   {\ensuremath{y_{\mathrm{cut}}}}

\newcommand{\PZz}  {\ensuremath{\mathrm{Z}}}
\newcommand{\PZzt} {Z}
\newcommand{\Pb}   {\ensuremath{\mathrm{b}}}
\newcommand{\Pbt}  {b}

\newcommand{\Plight}  {\ensuremath{\mathrm{dusc}}}

\newcommand{\bbbar}{\ensuremath{\mathrm{b\bar b}}}
\newcommand{\mZ}   {\ensuremath{m_{\mathrm{Z}}}}
\newcommand{\mb}   {\ensuremath{m_{\mathrm{b}}}}
\newcommand{\mbMS}   {\ensuremath{\overline{m}_{\mathrm{b}}}}
\newcommand{\MS}   {\ensuremath{\overline{\mathrm{MS}}}}

\newcommand{\ExtendedWeightedMean} {
  \mbMS(\mZ) = ( 2.67
  \pm 0.03
  {\mathrm{~(stat.)}}
  ^{+0.29}_{-0.37}
  {\mathrm{~(syst.)}}
  \pm 0.19
  {\mathrm{~(theo.)}} ) {\mathrm{~GeV}}}
\newcommand{\chiSquareN} {
  \mbMS(\mZ) = 2.67
  {\mathrm{~GeV}}}
\newcommand{\sevolvedToThreshold} {
  \mbMS(\mbMS) = ( 3.95
  ^{+0.52}_{-0.62} ) {\mathrm{~GeV}}}
\newcommand{\evolvedToThreshold} {
  \mbMS(\mbMS) = ( 3.95
  \pm 0.04 {\mathrm{~(stat.)}}
  ^{+0.43}_{-0.55} {\mathrm{~(syst.)}}
  \pm 0.28 {\mathrm{~(theo.)}} ) {\mathrm{~GeV}}}

\begin{document}
%
% \layout
%
%  Title Page:
%
\begin{titlepage}
\begin{center}{\large   EUROPEAN ORGANIZATION FOR NUCLEAR RESEARCH
}\end{center}\bigskip
\begin{flushright}
       CERN-EP-2001-034   \\ 4 May 2001
\end{flushright}
\bigskip\bigskip\bigskip\bigskip\bigskip
\begin{center}{\huge\bf   Determination of the \Pbt\ Quark Mass at the
    \PZzt\ Mass Scale
}\end{center}\bigskip\bigskip
\begin{center}{\LARGE The OPAL Collaboration
}\end{center}\bigskip\bigskip
\bigskip\begin{center}{\large  Abstract}\end{center}
\begin{sloppypar}
\noindent
In hadronic decays of \PZz\ bosons recorded with the OPAL detector at
LEP, events containing \Pb\ quarks were selected using the long lifetime of
\Pb\ flavoured hadrons. Comparing the $3$-jet rate in b 
events with that in d,u,s and c quark events, a significant difference
was observed.
Using ${\cal O}(\as^2)$ calculations for massive quarks, this
difference was used to determine the b quark mass in the
\ensuremath{\overline{\rm{MS}}} 
renormalisation scheme at the scale of the \PZz\ boson mass. By
combining the results from seven different jet finders
the running b quark mass was determined to be
\begin{displaymath}
 \ExtendedWeightedMean\;.
\end{displaymath}
Evolving this value to the b quark mass scale itself
yields
${ \sevolvedToThreshold }$,
consistent with results obtained at the b quark production
threshold. This determination confirms the QCD expectation of a scale
dependent quark mass. A constant mass is ruled out by 3.9 standard
deviations.
\end{sloppypar}
\bigskip\bigskip\bigskip\bigskip
\bigskip\bigskip
\begin{center}{\large
(Submitted to European Physical Journal C)
}\end{center}
%
%  Authors and time stamp:
%
%Authors: R. Seuster, O. Biebel\\
%
%Editorial Board: J.W. Gary, S. Kluth, A. De Roeck, T. Wengler
\end{titlepage}
%%%%%%%%%%%%%%%%%%%%%%%%%%%%%%%%%%%%%
\begin{center}{\Large        The OPAL Collaboration
}\end{center}\bigskip
\begin{center}{
%begin authorlist PLEASE DO NOT DELETE THIS COMMENT
G.\thinspace Abbiendi$^{  2}$,
C.\thinspace Ainsley$^{  5}$,
P.F.\thinspace {\AA}kesson$^{  3}$,
G.\thinspace Alexander$^{ 22}$,
J.\thinspace Allison$^{ 16}$,
G.\thinspace Anagnostou$^{  1}$,
K.J.\thinspace Anderson$^{  9}$,
S.\thinspace Arcelli$^{ 17}$,
S.\thinspace Asai$^{ 23}$,
D.\thinspace Axen$^{ 27}$
% ...
%end authorlist PLEASE DO NOT DELETE THIS COMMENT
}\end{center}\bigskip
\bigskip
%begin institutes
$^{  1}$School of Physics and Astronomy, University of Birmingham,
Birmingham B15 2TT, UK
\newline
$^{  2}$Dipartimento di Fisica dell' Universit\`a di Bologna and INFN,
I-40126 Bologna, Italy
\newline
$^{  3}$Physikalisches Institut, Universit\"at Bonn,
D-53115 Bonn, Germany
\newline
% ...
$^{  5}$Cavendish Laboratory, Cambridge CB3 0HE, UK
\newline
% ...
$^{  9}$Enrico Fermi Institute and Department of Physics,
University of Chicago, Chicago IL 60637, USA
\newline
% ...
$^{ 16}$Department of Physics, Schuster Laboratory, The University,
Manchester M13 9PL, UK
\newline
$^{ 17}$Department of Physics, University of Maryland,
College Park, MD 20742, USA
\newline
% ...
$^{ 22}$Department of Physics and Astronomy, Tel Aviv University,
Tel Aviv 69978, Israel
\newline
$^{ 23}$International Centre for Elementary Particle Physics and
Department of Physics, University of Tokyo, Tokyo 113-0033, and
Kobe University, Kobe 657-8501, Japan
\newline
% ...
$^{ 27}$University of British Columbia, Department of Physics,
Vancouver BC V6T 1Z1, Canada
\newline
% ...
Tel Aviv 69978, Israel.
%end notes
%%%%%%%%%%%%%%%%%%%%%%%%%%%%%%%%%%%%%
%
%  The main text:
%
\newpage
\section{Introduction}
\label{sec-introduction}
In Quantum Chromodynamics (QCD) the renormalisation group equation
(RGE) governs the energy dependence of both the renormalised coupling
$\as$ and the renormalised quark mass $m_{\rm q}$. 
The RGE for an observable $R$ calculated for massive quarks q and
measured at a  scale $Q$, states that $R$ is independent of the
renormalisation scale $\mu$ \cite{Ellis-Stirling-Webber}, which is
expressed by
\begin{equation}
\label{eqn-rge-mb}
\Big[ \mu^2 \frac{\partial}{\partial \mu^2}+
        \beta(\as) \frac{\partial}{\partial \as}-
        \gamma(\as) m_{\rm q} \frac{\partial}{\partial m_{\rm q}} \Big] 
        R({Q^2}/{\mu^2},\as,{m_{\rm q}}/{Q}) = 0 \; ,
\end{equation}
with the $\beta$ function $\beta(\as)$ of QCD and the mass anomalous
dimension $\gamma(\as)$.
This equation can be solved by introducing both a running coupling
constant $\as(Q^2)$ and a running quark mass $m_{\rm q}(Q^2)$.
In particular, 
the scale dependence of the b quark mass in the
\ensuremath{\overline{\rm{MS}}} renormalisation scheme, $\mbMS(Q^2)$, is
to four-loop accuracy given by
\begin{eqnarray}
\label{eqn-running-mb}
 \mbMS(Q^2) & = &  \nonumber
  \hat{m}_{\rm b} \cdot 
  \left(\frac{\as(Q^2)}{\pi}\right)^{12/23}
 \cdot \\ \nonumber
 & &
  \left[1 + 1.175\cdot\left(\frac{\as(Q^2)}{\pi}\right) +
        1.500\cdot\left(\frac{\as(Q^2)}{\pi}\right)^2 + \right. \\
 & & \left. \hspace*{1.0cm} 0.172\cdot\left(\frac{\as(Q^2)}{\pi}\right)^3
        + {\cal{O}} \left(\as(Q^2)^4\right) \right]
\quad ,
\end{eqnarray}
taking the renormalisation group invariant\footnote{i.e. independent of
$\mu$} mass $\hat{m}_{\rm b}$ as a reference, 
see e.g. \cite{bib-Vermaseren-PLB405-327}. Analogous to $\as(Q^2)$, an
absolute value for $\mbMS(Q^2)$ is not predicted by QCD.
A b quark mass of 4.2 GeV \cite{RPP2000} 
measured at the production threshold
corresponds to a running mass $\mbMS(Q^2)$ 
of about 3 GeV in interactions at the scale of the $\PZz$ mass.
The experimental observation of this running of the quark mass
constitutes an important test of QCD.

Studies of the flavour dependence of the strong coupling constant
observed a difference in jet rates and event shapes between b events
and light quark events, see
e.g. \cite{bib-alphas-flavour}. This apparent deviation of a few
percent from a flavour-independent coupling constant can be explained
by effects of the large b quark mass. Second order matrix elements
that have been calculated recently, taking finite quark masses fully into
account \cite{bib-Bernreuther-PRL79-189,
  bib-Rodrigo-PRL79-193,bib-Nason-NPB521-237} can explain these
experimental observations. Flavour independence of the strong
interaction is a fundamental property of QCD. Assuming it holds, the
second order matrix elements for massive
quarks can be used to determine the b quark mass at scales different
from production threshold.

For the determination of the running b quark mass, the ratio of 3-jet
rates in b events, $R_3^{\rm b}$, over 3-jet rates in light quark events,
$R_3^{\rm dusc}$,
\begin{equation}
\label{eqn-ratio-B3}
 B_3 = \frac{R_3^{\rm b}}{R_3^{\rm dusc}}
\end{equation}
has been proposed in \cite{bib-Bernreuther-PRL79-189}.
This ratio
is sensitive to mass dependent differences in gluon radiation from b
and from light quarks. 
This or equivalent methods have been used in determinations of the b
quark mass by DELPHI \cite{bib-mb-at-mZ}, ALEPH \cite{ALEPH_mb}
 and Brandenburg et al. using SLD data
\cite{bib-SLAC-PUB-7915}.

In this paper we present a determination of the running b quark mass
based on the variable $B_3$ using the large statistics sample
collected with the OPAL detector  
 at the ${\rm e}^+{\rm e}^-$ collider LEP at centre-of-mass
energies close to the \PZz\ mass.
Events containing b hadrons were tagged by
identifying their displaced decay vertices. The light and the b quark
contributions were deduced from the tagged and inclusive samples by a
simple unfolding technique which relies only on the tagging
efficiencies and fake tagging rates which were estimated by studying
Monte Carlo events.

\section{The OPAL detector, data, and Monte Carlo simulation}
\label{sec-detector}
A detailed description of the OPAL detector can be found elsewhere
\cite{bib-OPALdet}. For the present analysis only ${\rm e}^+{\rm e}^-$
collisions collected in 1994 were included, as these
provide sufficient statistics
with a uniform detector configuration.
The silicon strip micro-vertex detector, all
central tracking detectors and the electro-magnetic calorimeter were
required to be fully operational. 
Standard criteria for high multiplicity hadronic events
\cite{bib-GCE+HIMS,bib-TKMH}, which rely on a minimum number of
measured tracks in the central tracking system and clusters in
the electro-magnetic lead glass calorimeter were applied. The 
remaining background, mostly from two-photon processes and $\tau$-pair
events, was estimated to be $0.5\,\%$ and $0.11\,\%$,
respectively \cite{bib-TKMH}.
\par
The silicon micro-vertex detector 
\cite{bib-SIdetector}
is used for the identification of b quark events based on
lifetime information.
To account for the limited polar angle acceptance of the silicon
micro-vertex detector
operating in 1994, only events whose thrust vector pointed to the
central part of the detector, $|\cos\theta_{\rm Thrust}|<0.75$, were
considered\footnote{OPAL uses a right handed coordinate system with
the $z$ axis pointing along the electron beam direction and $x$
towards the centre of the LEP ring. The polar angle
$\theta$ is measured with respect to the $z$ axis.
}.
After this cut about $924\thinspace 000$ events remained. These
events defined the inclusive sample.
Tracks of charged particles recorded in the tracking detectors and
clusters of energy recorded in the calorimeters were used 
for jet finding and were required
to satisfy a set of standard quality cuts which are detailed in
\cite{bib-OPAL-ZPC72-191}.
The energy of each cluster associated to a track was corrected for
double counting of energy using the momentum of that track
\cite{bib-OPAL-EPJC2-213}.

To determine the efficiency and purity of the event selection and to 
correct for distortions due to the finite acceptance and resolution of 
the detector, about $4$ million hadronic decays of the $\PZz$ were
generated by the JETSET program version 7.4 \cite{bib-JETSET},
tuned to describe OPAL data \cite{bib-OPALtune}. The
generated events were passed through a detailed simulation of the OPAL
detector \cite{bib-GOPAL} and reconstructed using the same procedures
as for the data. The b quark events in the Monte Carlo sample were
reweighted to correspond to the most 
recent estimates for the
parameter of the Peterson et al. fragmentation function
\cite{bib-Peterson-fragmentation}, $\epsilon_{\rm b}$ \cite{bib-OPALtune},
the mean charged particle decay multiplicity in b events,
$n_{\rm charged}^{\rm decay}$ \cite{LEPEWWG-n_ch}, and the mean b hadron
lifetime, $\tau_{\rm B}$ \cite{LEPEWWG-tau}. The c quark events were
reweighted to correspond to a recent estimate of the Peterson
fragmentation parameter, $\epsilon_{\rm c}$ \cite{bib-OPALtune}. The
reweighting procedures are described in \cite{bib-OPAL_RB}.

\section{Selection of b quark events}
\label{sec-selection}
The silicon micro-vertex detector was used in addition to the
tracking detectors to
measure the decay length of b flavoured hadrons in hadronic \PZz\
decays. The decay length is defined by the distance between the
reconstructed primary vertex and the identified b hadron decay vertex.
These vertices were reconstructed using the algorithm
described in \cite{bib-OPAL-ZPC65-17}. In the procedure 
a cone jet algorithm \cite{OPAL-PR097} is applied to search for jets
in each event, using a cone half angle with $R=0.55~\mathrm{rad}$ and
a minimum jet energy of $5.0~ \mathrm{GeV}$, as in
\cite{bib-OPAL_RB}. A common secondary vertex was searched for
in such a jet by iteratively excluding the track with the largest
$\chi^2$ contribution and repeating the fit until all $\chi^2$
contributions were smaller than 4. A minimum number of three tracks
was required to form a vertex. For each event the vertex with the
largest decay length significance $L/\sigma_L$ was determined, where
$L$ is the decay length and $\sigma_L$ its uncertainty.
\par
\begin{figure}
\begin{center}
  \mbox{\includegraphics[width=1.0\linewidth]{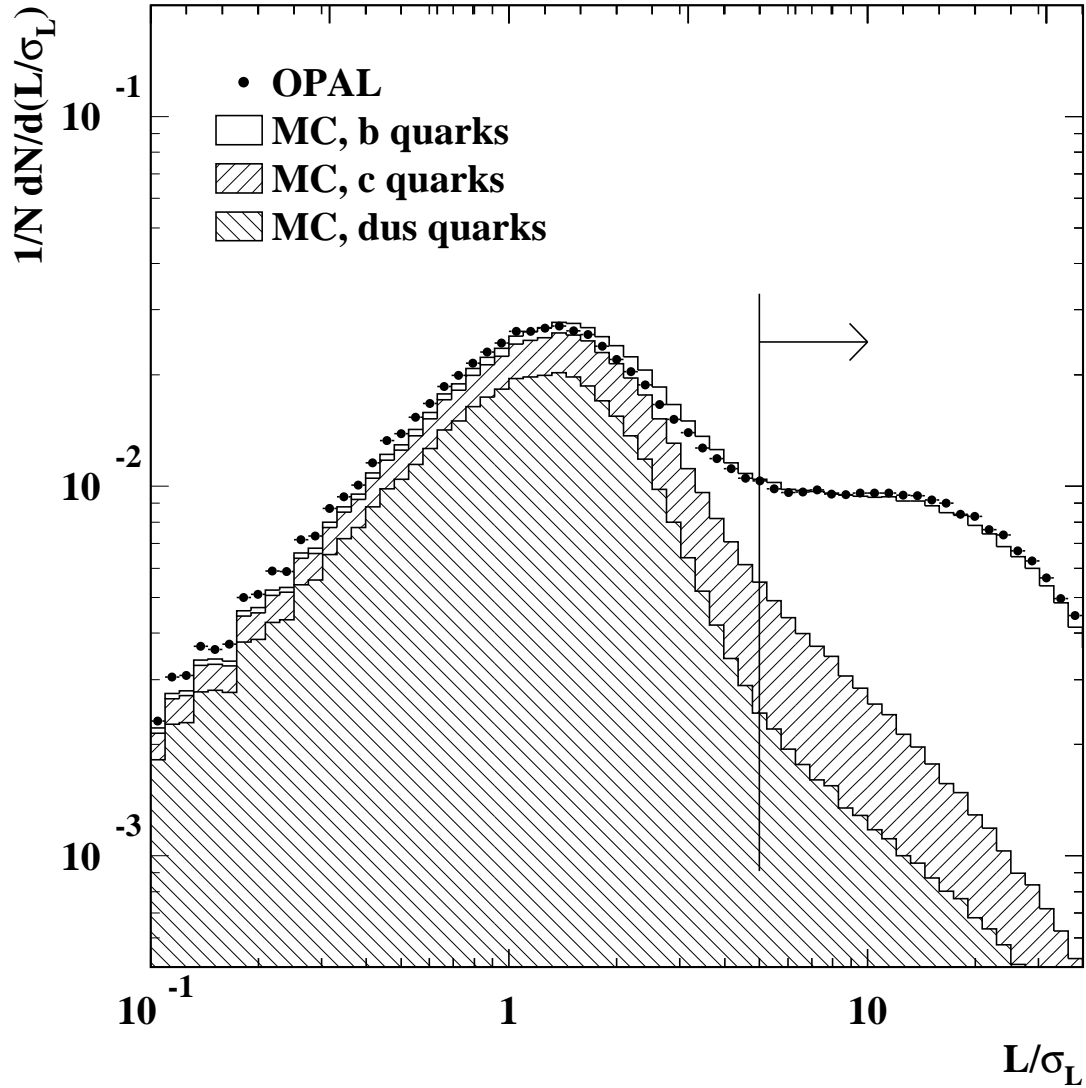}}
\end{center}
\caption{\label{fig-decaylength-significance}
  Distribution of the decay length significance $L/\sigma_L$ for data 
  (points) and simulation (histograms). The contributions from the 
  d, u or s and c quark events in the simulation
  are shown by different hatching, b quark events are shown by
  the open histogram. The vertical line indicates 
  the cut chosen to select b quark events. The
  statistical uncertainty is also shown.
  }
\end{figure}

The distribution of $L/\sigma_L$ is shown
in Figure \ref{fig-decaylength-significance}.
A good agreement between data and simulation is observed in
the region $L/\sigma_L > 5$, used to select a sample enriched in b
events.
In the Monte
Carlo \Pb\ quark events are selected with an efficiency of $\epsilon =
(70.23 \pm 0.02) \%$, whereas the fake tag rate from light quark
events being mis-identified as a \Pb\ candidate is $f = (7.20 \pm
0.01) \%$, where the uncertainties are statistical only. $21.3\%$
of all hadronic events are tagged as \Pb\ candidates in the data,
compared with $21.0\%$ in the simulation. This small deviation will
be discussed in section \ref{det_and_b_tag}.

\section{Measurement of $B_3$}
\label{sec-B3ratio}
For the determination of $B_3$
jets were reconstructed using the standard JADE algorithm, its
variants E, E0, P and P0 and the DURHAM and the GENEVA algorithms, all
described in 
\cite{bib-Bartel-ZP33-23} and \cite{bib-Bethke-NPB370-310}.
In addition, the CAMBRIDGE algorithm\cite{Cambridge} was
used. However, since there is not yet a second order calculation
compatible with our definition of $B_3$ for this jet finder available,
it was not used to determine the b quark mass.
All these jet finders combine the two objects with the smallest
distance as measured using the distance measures in Table
\ref{tab_jetfinder}. These two objects are combined
according to the prescription given in the third column of Table
\ref{tab_jetfinder} to form a new object. This procedure is
repeated until all distances between objects are
larger than a resolution parameter \yc. The number of jets is then
given by the number of remaining objects.
To limit the uncertainty related to the choice of a specific jet
finder, seven different jet finders were used in this analysis.

\begin{table}
\begin{center}
\begin{tabular}[h]{|c||c|c|}
\hline
algorithm  & distance measure $y_{ij}$&       recombination \\ \hline \hline

JADE    & $2 E_i E_j (1-\cos \theta_{ij} ) / s $ & $ p_k=p_i+p_j $ \\ \hline
JADE E0 & $(p_i+p_j)^2 / s $                     & $ E_k=E_i+E_j;$ 
        $ {\bf p}_k = E_k \frac{({\bf p}_i+{\bf p}_j)}{|{\bf p}_i+{\bf p}_j|}$\\ \hline
JADE E  & $(p_i+p_j)^2 / s $                     & $ p_k=p_i+p_j $ \\ \hline
JADE P  & $(p_i+p_j)^2 / s $                     & $ {\bf p}_k={\bf p}_i+{\bf p}_j; $
        $ E_k = |{\bf p}_k| $ \\ \hline
JADE P0 & $(p_i+p_j)^2 / (\sum_l E_l)^2 $        & $ {\bf p}_k={\bf p}_i+{\bf p}_j; $
        $ E_k = |{\bf p}_k| $ \\ \hline \hline
DURHAM  & $2 {\rm min}(E_i^2, E_j^2) (1- \cos \theta_{ij}) / s $ & $p_k=p_i+p_j$ \\ 
  \hline
GENEVA  & $8 E_i E_j (1 - \cos \theta_{ij}) / 9(E_i + E_j)^2$ & $p_k=p_i+p_j$ \\
  \hline
CAMBRIDGE & $2 (1 - \cos \theta_{ij})$ & $p_k=p_i+p_j$ \\
   & soft freezing if $y_{ij} > y_{\rm cut}$ & \\ \hline
\end{tabular}
\end{center}
\caption{\label{tab_jetfinder}
  Definitions of the distance measures and recombination prescriptions
  for the different jet finders used in this analysis
  \cite{bib-Bartel-ZP33-23,bib-Bethke-NPB370-310}.
  $p_i$, ${\bf p}_i$, $E_i$ describe the four-, the three-momenta and
  the energy of particle $i$. $\theta_{ij}$ is the angle between the
  three-momenta of particles $i$ and $j$. The
  recombination prescription of the CAMBRIDGE jet finder is described
  in more detail in \cite{Cambridge}.}
\end{table}
The double ratio $B_3(\yc)$ was determined from the $3$-jet rates in
the event sample enriched in b quarks and in the inclusive event
sample at values for $\yc$ at which the predictions were calculated,
see Table \ref{tab-correction}.
As the contribution of events not originating from hadronic \PZz\
decays is very small, the inclusive sample can be decomposed
into a b quark sample plus a light quark sample.
The number of events, ${\cal A}$, in the inclusive sample and the
number of events, ${\cal T}$, in the b enriched sample can be written
in terms of the number of b quarks events, $N^{\Pb}$, and of
light quark events, $N^{\Plight}$:
\begin{eqnarray}
{\cal A} & = & N^{\Pb} + N^{\Plight} \\
{\cal T} & = & \epsilon N^{\Pb} + f N^{\Plight}
\end{eqnarray}
where $\epsilon$ is the Monte Carlo tagging efficiency for
the b quark events and $f$ is the fake tag rate for light quark
events.
These two equations can be solved for $N^{\Pb}$ and $N^{\Plight}$.
A similar decomposition is valid for the number of 3-jet events,
$a_3(\yc)$, in the inclusive sample and the number of 3-jet events,
$t_3(\yc)$, in the b enriched sample, with the number of $3$-jet
b quark events, $n^{\Pb}_3(\yc)$, and the number of $3$-jet light quark
events, $n^{\Plight}_3(\yc)$:
\begin{eqnarray}
a_3(\yc) & = & n^{\Pb}_3(\yc) + n^{\Plight}_3(\yc) \\
t_3(\yc) & = & \epsilon_3(\yc) n^{\Pb}_3(\yc) + f_3(\yc) n^{\Plight}_3(\yc)
\end{eqnarray}
The variables $\epsilon_3$ and $f_3$, which depend on \yc, are the
corresponding efficiency and fake tagging rate for 3-jet events.
Solving these last two equations for $n^{\Pb}_3(\yc)$ and
$n^{\Plight}_3(\yc)$ and using the similar equations for 
$N^{\Pb}$ and $N^{\Plight}$ leads to the following relation:
\begin{eqnarray}
\label{eqn-B3-determination}
  B_3(\yc) & = & 
    {\cal C}_{\mathrm{had}}(\yc) \cdot
    {\cal C}_{\mathrm{det}}(\yc) \cdot
    \frac{R_3^{\Pb}(\yc)}{R_3^{\Plight}(\yc)}
 \nonumber \\
        & = &
    {\cal C}_{\mathrm{had}}(\yc) \cdot
    {\cal C}_{\mathrm{det}}(\yc) \cdot
    \frac{n^{\Pb}_3(\yc)/N^{\Pb}}{n^{\Plight}_3(\yc)/N^{\Plight}}
 \nonumber \\
        & = &
    {\cal C}_{\mathrm{had}}(\yc) \cdot
    {\cal C}_{\mathrm{det}}(\yc) \cdot \nonumber \\
        & & 
    \frac{t_3(\yc)-f_3(\yc) \cdot a_3(\yc)}
         {{\cal T} -f \cdot {\cal A}} \cdot
    \frac{{\cal T} - \epsilon \cdot {\cal A}}
         {t_3(\yc) - \epsilon_3(\yc) \cdot a_3(\yc)}
\quad .
\end{eqnarray}
Since $B_3$ is a ratio,
common correction
factors for the individual 3-jet rates of b quark and light quark
events cancel.
We apply bin-by-bin correction factors for detector distortions,
${\cal C}_{\mathrm{det}}(\yc)$, and hadronisation effects,
${\cal C}_{\mathrm{had}}(\yc)$, as shown in
Eq. (\ref{eqn-B3-determination}).
The correction factors are defined as the ratio of the double ratio of
the 3-jet rates for b over light quark events
determined from the simulation at either the hadron or parton level
divided by the same double ratio at detector or hadron level.
The hadron level consists of particles generated
by the Monte Carlo program with a mean lifetime greater than 300
ps. The partons which are present at the end of the parton shower in
the generator define the parton level. The parton shower in JETSET,
which we used to estimate the size of the hadronisation corrections,
terminates when partons reach virtualities below a cut-off
$Q_0$, set to $1.9~\mathrm{GeV}$ in the standard analysis.
\par
\begin{figure}
\begin{center}
  \mbox{\includegraphics[width=1.0\linewidth]{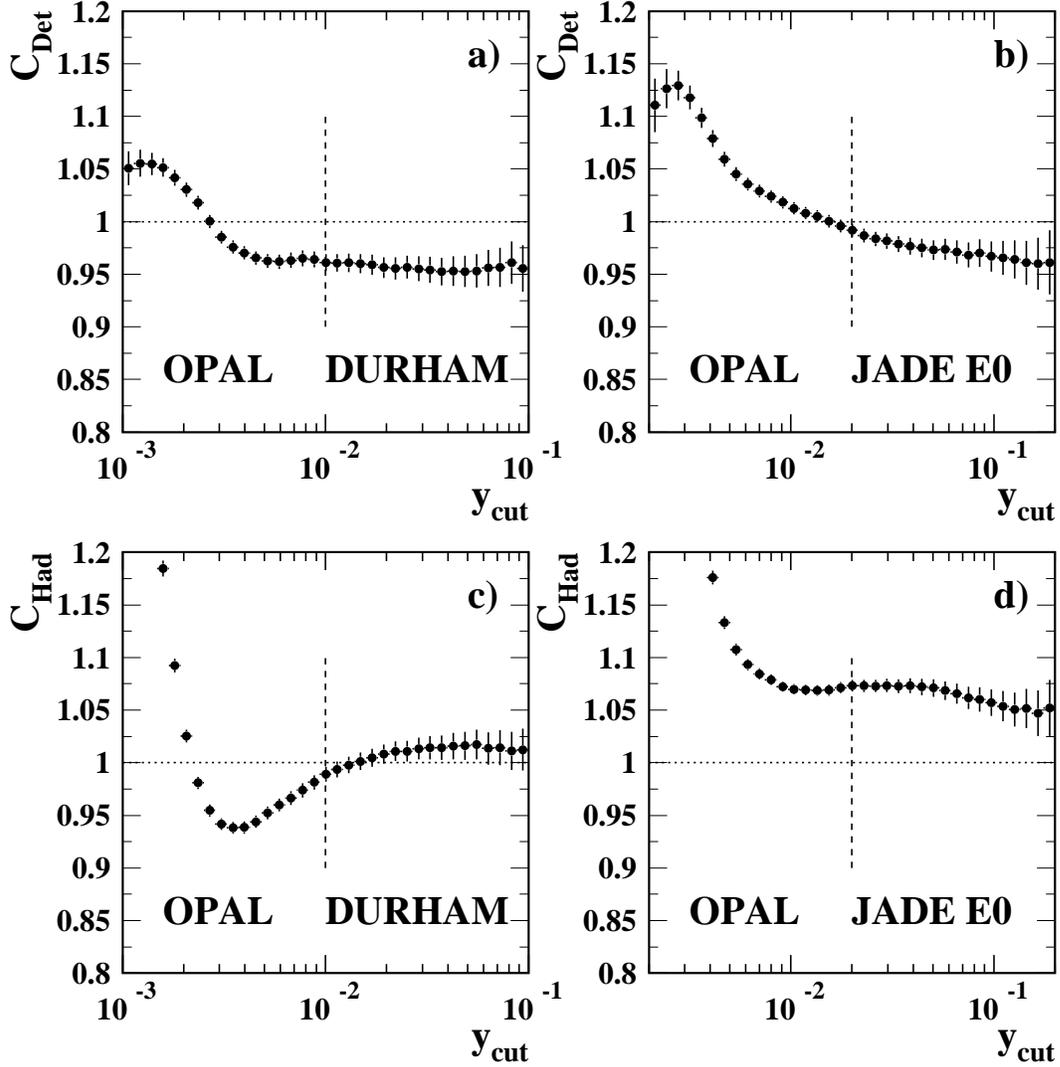}}
\end{center}
\caption{\label{fig-correction}
  In a) and b) the correction factors of the $B_3$ ratio for detector
  effects, ${\cal C}_{\mathrm{det}}$, and in c) and d) for
  hadronisation effects, ${\cal C}_{\mathrm{had}}$, are shown as a
  function of $\yc$ for the DURHAM (left column) and JADE E0 (right
  column) jet finders. The error bars show the statistical
  uncertainty of the Monte Carlo. The dashed vertical lines
  indicate the $\yc$ values chosen to determine the b quark
  mass.
  }
\end{figure}

Figure \ref{fig-correction} shows the $\yc$ dependence of the 
correction factors for the DURHAM and the JADE E0 algorithms. The
DURHAM scheme has the smallest, the JADE E0 scheme the largest
hadronisation corrections of all schemes used in this analysis. The
correction factors for the other jet finders are summarised in Table
\ref{tab-correction}.
\begin{table}
\begin{center}
\begin{tabular}{|l||c||c|c|}
\cline{2-4}
\multicolumn{1}{c|}{ }
& $\yc$ & $\mathrm{\cal C}_{\mathrm{Det}}$ & $\mathrm{\cal C}_{\mathrm{Had}}$
\\ \hline
JADE
 & 0.02
 & 0.965
 & 1.016
\\ \hline
DURHAM
 & 0.01
 & 0.961
 & 0.989
\\ \hline
JADE E0
 & 0.02
 & 0.993
 & 1.073
\\ \hline
JADE P
 & 0.02
 & 0.978
 & 1.015
\\ \hline
JADE P0
 & 0.015
 & 0.985
 & 1.022
\\ \hline
JADE E
 & 0.04
 & 0.997
 & 1.049
\\ \hline
GENEVA
 & 0.08
 & 0.971
 & 1.031
\\ \hline
CAMBRIDGE
 & 0.01
 & 0.971
 & 1.017
\\ \hline
\end{tabular}
\caption{\label{tab-correction}
  Correction factors for each jet finder for detector distortions and
  hadronisation effects. The JETSET Monte Carlo program was used to
  estimate the size of the corrections.}
\end{center}
\end{table}

\begin{figure}
\begin{center}
  \mbox{\includegraphics[width=1.0\linewidth]{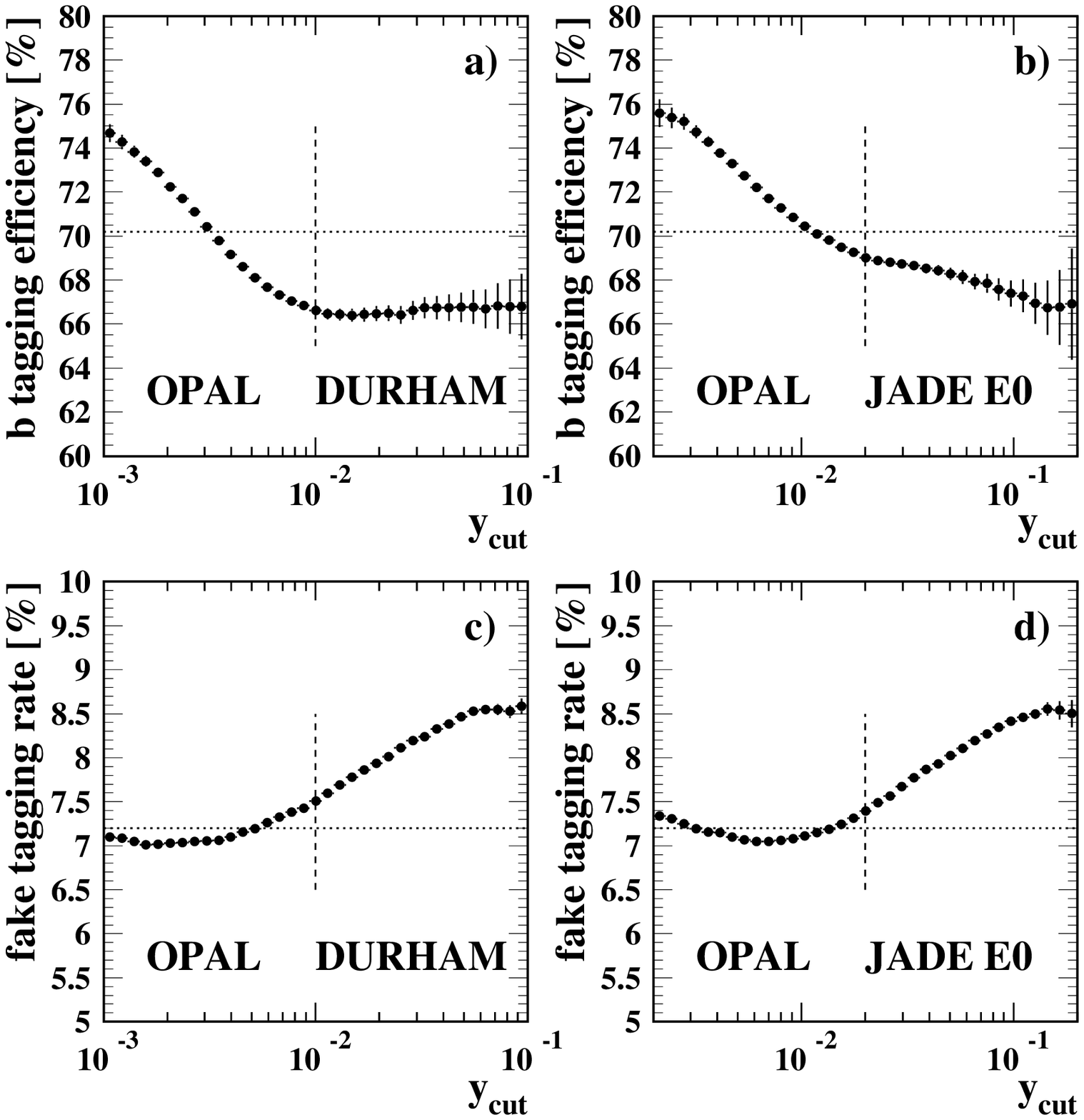}}
\end{center}
\caption{\label{fig-efficiency}
  a) and b) show the efficiency $\epsilon_3(\yc)$ for b quark 3-jet
  events for the DURHAM
  (left) and JADE E0 (right column) jet finders. Overlaid as a dotted
  line is $\epsilon$, the tagging efficiency for all b quark events; c)
  and d) show the fake tag rates $f_3(\yc)$ for light quark 3-jet
  events for both jet finders. Overlaid is the fake tag rate $f$ for
  all light quark events. The error bars describe the statistical
  uncertainty of the Monte Carlo.
  The dashed vertical lines indicate the $\yc$ values chosen to
  determine the b quark mass.
  }
\end{figure}

The efficiency and the fake tag rate for 3-jet events in b or light
quark events, respectively, shown for the DURHAM and
JADE E0 jet finder in Figure \ref{fig-efficiency}, depend slightly on
the chosen $\yc$ value because of the difference in kinematics induced
by the presence of a highly energetic gluon emitted at large angle.

\begin{figure}
\begin{center}
  \mbox{\includegraphics[width=1.0\linewidth]{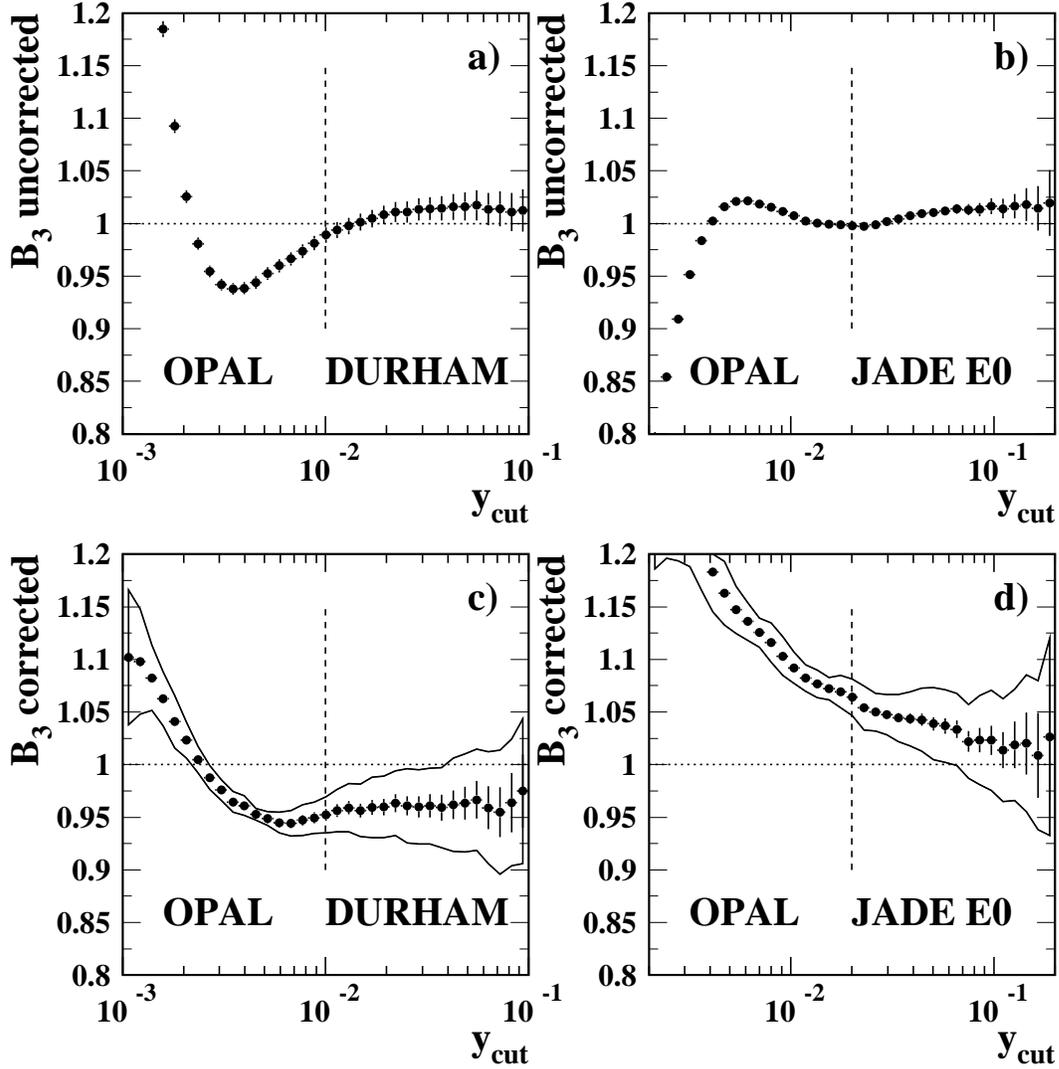}}
\end{center}
\caption{\label{fig-corrected-B3}
  The uncorrected a) and b), and the corrected c) and d) $B_3$
  ratios for the DURHAM and the JADE E0 jet
  finders. The uncertainties include the effect of
  finite statistics in both data and Monte Carlo. There is a
  substantial correlation between adjacent bins. The dashed vertical
  lines indicate the $\yc$ values chosen for the determination of
  the b quark mass. The bands in c) and d) display the total
  systematic and the statistical uncertainty added in quadrature.
  }
\end{figure}

In Figure \ref{fig-corrected-B3} the measured $B_3$ ratio is shown
both before applying corrections and after being corrected to the
parton level.
The detector correction factors, which take into account kinematic
biases induced by the b tagging, are usually larger than the
hadronisation correction, which includes known decays of
b flavoured hadrons in the b quark sample. Only for JADE E0 and 
JADE E these corrections are larger than the detector
corrections. All other hadronisation correction factors are smaller
than about $3\%$.
\newpage
\section{Determination of $\mbMS$}
To determine the \Pb\ quark mass from the measured $B_3$ ratio,
parametrisations of the QCD predictions for the $3$-jet rate for b and
light quarks were used. For \Pb\ quarks the
predictions\cite{bib-Bernreuther-PRL79-189,Brandenburg}
 cover masses in the range from $2$ to $4$~GeV in steps of
$0.2$~GeV at one specific $\yc$ value for each jet finder. These $\yc$
values are 
listed in Table \ref{tab-correction}. For light quarks the
parametrisation of \cite{bib-Bethke-NPB370-310} was adopted
except for the DURHAM jet finder, for which a more recent
parametrisation from \cite{Cambridge} was used.

Using these parametrisations the double ratio 
$B_3=R_3^{\rm b}/R_3^{\rm dusc}$ was derived:
\begin{eqnarray}
B_3(\yc,\overline{m}_{\mathrm{b}}) & = &
 \frac{R_3^{\rm b}(\yc,\overline{m}_{\mathrm{b}})}{R_3^{\rm dusc}(\yc)} \nonumber \\
 & = &
 \frac{
  \left(\frac{\as}{2\pi}\right)   A^{\rm b}(\yc,\overline{m}_{\mathrm{b}}) +
  \left(\frac{\as}{2\pi}\right)^2 B^{\rm b}(\yc,\overline{m}_{\mathrm{b}})
}{
  \left(\frac{\as}{2\pi}\right)   A(\yc) + 
  \left(\frac{\as}{2\pi}\right)^2 [B(\yc) -2\cdot A(\yc) ]
} \; ,
\end{eqnarray}
where the coefficients 
$A^{\rm b}(\yc,\overline{m}_{\mathrm{b}})$ and
$B^{\rm b}(\yc,\overline{m}_{\mathrm{b}})$ and
$A(\yc)$ and $B(\yc)$ parametrise the 3-jet rates for massive and
massless quarks. 
Note that 
the $B$ coefficients for massive and for massless
  quarks differ in their definitions because the 3-jet rate for massive
  quarks is normalised to the total $\rm b\bar{b}(g)$ cross section, whereas
  the 3-jet rate for massless quarks is normalised to the hadronic
  born, i.e. only $\rm q\bar{q}$, cross section.
To obtain a
prediction in a finite order of $\as$, the denominator of the double
ratio was expanded in a Taylor series after cancelling one order of
$\as$ in numerator and denominator.
Due to this cancellation of one order of $\as$, the QCD predictions in
this expanded expression for $B_3$ are of ${\cal O}(\as)$. Due to the
additional dependence of $B_3$ on the b quark mass, the dependence on
the renormalisation scale $x_{\mu}=\mu/Q$ enters in first order of
$\as$, as can be seen from the expanded expression:
\begin{eqnarray}
\label{variation_mu}
B_3(\yc,\overline{m}_{\mathrm{b}},x_{\mu}) & = &
\frac{A^{\rm b}(\yc,\overline{m}_{\mathrm{b}})}{A(\yc)}+
\frac{\as}{2\pi} \cdot \nonumber  \\ & & \left\{
\frac{B^{\rm b}(\yc,\overline{m}_{\mathrm{b}})}{A(\yc)}-
\frac{A^{\rm b}(\yc,\overline{m}_{\mathrm{b}})}{A(\yc)} \cdot
\frac{B(\yc)-2 \cdot A(\yc)}{A(\yc)}+
 \nonumber \right. \\ & & \left.
2\pi \gamma_0\overline{m}_{\mathrm{b}} \cdot
\frac{\partial A^{\rm b}(\yc,\overline{m}_{\mathrm{b}}) \; / \partial 
\overline{m}_{\mathrm{b}}}{A(\yc)}
\log x_{\mu}^2 \right\},
\end{eqnarray}
where $\gamma_0$ is the first coefficient in the perturbative
expansion of the anomalous mass dimension.
The renormalisation scale parameter $x_{\mu}$ was
set to unity at the renormalisation scale of 91.2 GeV and
$\as$ was set to its world average of 0.1184 \cite{bib-Bethke-0004021}.
To parametrise the mass dependence of Eq.(\ref{variation_mu}) in the
range from 2 to 4 GeV a parabolic function
\begin{equation}
    B_3^i(\mbMS) = a_0^{i} + a_2^{i}\cdot \mbMS^2 \quad, \quad
    i=\mathrm{JADE},\mathrm{E},\mathrm{E0},\mathrm{P},\mathrm{P0},
    \mathrm{DURHAM},\mathrm{GENEVA}
\label{b3_2_mb}
\end{equation}
was used, where the coefficients and their uncertainty due to a finite
Monte Carlo integration sample for $R_3^{\Pb}$ are given in Table
\ref{tab-coefficients}. The last column gives the
$\chi^2/\mathrm{d.o.f.}$ of the fits.
\begin{table}
\begin{center}

\begin{tabular}{|l||c||c|c||c|c||c|}
\cline{2-7}
\multicolumn{1}{c|}{ }
 & $\yc$ & $a_0^i$ & $\Delta a_0^i$ &
   $a_2^i \times 1000$ & $\Delta a_2^i \times 1000$
 & $\chi^2/\mathrm{d.o.f.}$
 \\ \hline
 JADE & 0.02
 & 0.9909
 & 0.0013
 & -2.505
 & 0.125
 & 0.233 
 \\ \hline
 DURHAM & 0.01
 & 0.9918
 & 0.0025
 & -3.771
 & 0.230
 & 0.383 
 \\ \hline
 JADE E0 & 0.02
 & 1.0142
 & 0.0013
 &  6.093
 & 0.124
 & 3.283 
 \\ \hline
 JADE P & 0.02
 & 0.9824
 & 0.0013
 &  3.845
 & 0.126
 & 1.580 
 \\ \hline
 JADE P0 & 0.015
 & 0.9756
 & 0.0021
 &  5.749
 & 0.186
 & 0.868 
 \\ \hline
 JADE E & 0.04
 & 1.0043
 & 0.0021
 &  9.886
 & 0.211
 & 0.979 
 \\ \hline
 GENEVA & 0.08
 & 1.0171
 & 0.0013
 & -2.406
 & 0.132
& 0.364
\\ \hline
\end{tabular}
\caption{\label{tab-coefficients}
  Coefficients for Eq. (\ref{b3_2_mb}) for the different jet finders.}
\end{center}
\end{table}
In \cite{bib-SLAC-PUB-7915} a different parametrisation was
chosen.
The difference between this parametrisation and the one we used in the
relevant mass region of 2 to 4 GeV is smaller than the statistical
uncertainty of the Monte Carlo integration.
To calculate the derivative 
$\partial A^{\rm b}(\yc,\overline{m}_{\mathrm{b}}) / \partial
\overline{m}_{\mathrm{b}}$ in Eq.
(\ref{variation_mu}) the same parametrisation as in
Eq. (\ref{b3_2_mb}) was fitted to the
$A^{\rm b}(\yc,\overline{m}_{\mathrm{b}})$ coefficient and the derivative
was determined.
\par
In \mbox{Table \ref{tab_results_jade}} the measured
$B_3$ values are shown, together with the results for
$\overline{m}_{\mathrm{b}}$ obtained using Eq. (\ref{b3_2_mb}) along
with their uncertainties, described in more detail in the next
section.
Correlations between the different event samples entering the
unfolding in Eq. (\ref{eqn-B3-determination}) were taken into
account.
These correlations were calculated from the full Monte
Carlo sample.
\begin{table}[h]
\begin{center}
\begin{tabular}{|l||c|c|c|}
\cline{2-4}
\multicolumn{1}{c|}{ }
& $\yc$ & $B_3$ & $\overline{m}_{\mathrm{b}} [\mathrm{GeV}]$ \\ \hline
 JADE
 & 0.02
 & $ 0.9752 \pm 0.0048 \pm 0.0129 $
 & $2.51 \pm 0.42 \pm 0.99 \pm 0.16$
\\ \hline
 DURHAM
 & 0.01
 & $ 0.9532 \pm 0.0056 \pm 0.0166 $
 & $3.20 \pm 0.24 \pm 0.64 \pm 0.18$
\\ \hline
 JADE E0
 & 0.02
 & $ 1.0657 \pm 0.0035 \pm 0.0164 $
 & $2.91 \pm 0.10 \pm 0.50 \pm 0.26$
\\ \hline
 JADE P
 & 0.02
 & $ 0.9938 \pm 0.0040 \pm 0.0152 $
 & $1.73 \pm 0.34 \pm 1.93 \pm 0.11$
\\ \hline
 JADE P0
 & 0.015
 & $ 1.0140 \pm 0.0035 \pm 0.0129 $
 & $2.58 \pm 0.12 \pm 0.45 \pm 0.09$
\\ \hline
 JADE E
 & 0.04
 & $ 1.0649 \pm 0.0032 \pm 0.0154 $
 & $2.47 \pm 0.07 \pm 0.32 \pm 0.25$
\\ \hline
 GENEVA
 & 0.08
 & $ 0.9817 \pm 0.0094 \pm 0.0257 $
 & $3.84 \pm 0.55 \pm 1.27 \pm 0.12$
\\ \hline
 CAMBRIDGE
 & 0.01
 & $ 0.9717 \pm 0.0062 \pm 0.0192 $
 & -
\\ \hline
\end{tabular}
\end{center}
\caption{\label{tab_results_jade}
  For each jet finder the double ratio $B_3$ and the corresponding
  value for the b quark mass are given. For the double ratio
  $B_3$ the statistical and systematic uncertainties are given.
  For the b quark mass 
  $\overline{m}_{\mathrm{b}}$ the uncertainties listed are the
  statistical, the total systematic and the total theoretical
  uncertainties.
}
\end{table}

\section{Systematic and theoretical uncertainties}
\label{sec-systematics}
Various sources of systematic uncertainty were investigated to 
assess their impact on the measured b quark mass.
Selection criteria and parameter values in the
Monte Carlo simulation were changed from their defaults and the
entire analysis was repeated. The deviation of the mass value from the
standard result was taken as a systematic uncertainty.

The investigations can be grouped into three classes according to the 
corrections and efficiencies in Eq. (\ref{eqn-B3-determination}).
These classes are either related to (i) detector and b tagging, (ii)
the hadronisation correction uncertainty for the Monte Carlo generator,
or (iii) theoretical uncertainty.
For any systematic variation which has a positive and a negative
contribution the larger of both was taken as the symmetric
uncertainty.
For the variations which have only one deviation, the varied cut on
the decay length significance, the detector simulation and the
modelling of the b quark fragmentation, the deviation was taken
as the symmetric uncertainty.
\par
Tables \ref{tab-systematics} and \ref{tab-translation} summarise the
results of these checks along with their uncertainties assigned.
The fairly large spread between the numerical values for the
systematic variations for the different jet finders is caused by
different slopes for the parametrisation of the double ratio in terms
of the b quark mass, Eq. (\ref{b3_2_mb}).

\subsection{Detector simulation and b-tagging uncertainties}
\label{det_and_b_tag}
Biases affecting the jet reconstruction due to the modelling of
tracks and clusters were estimated as follows.
To assess the uncertainty related to the simulation of tracks, the
resolution of reconstructed track parameters in the Monte Carlo was
changed by $\pm10\%$ \cite{bib-OPAL_RB} and the analysis repeated.
Similarly, to assess the uncertainty related to the simulation of the
electromagnetic calorimeter, the resolution of this detector was
changed by $\pm10\%$ in the Monte Carlo and the analysis repeated.
The largest deviation observed was taken as the uncertainty due to
detector simulation.
For all jet finders changing resolution of the track parameters by
$+10\%$, i.e. degrading the resolution, gave the largest deviation.
\par
The effect of the cut on the limited polar angular acceptance of the
silicon micro-vertex detector was estimated by changing  
$|\cos\theta_{\mathrm{Thrust}}|<0.75$ by $\pm 0.05$.
\par
To assess the impact of the \Pb\ quark selection cut, the
analysis was repeated requiring for the decay length significance 
$L/\sigma_L$ a minimum value of $8$ instead of $5$ in the standard
analysis, which reduced the efficiency to $(60.41 \pm 0.02)\,\%$ and
lowered the fake tag rate to $(4.77 \pm 0.01)\,\%$.
\par
The b tagging efficiency also depends on the mean number of charged
particles from b hadron decays, their mean lifetime and the
branching fraction of \PZz~ into $\mathrm{b\bar{b}}$, which were
varied within the range given in Tables \ref{tab-systematics} and
\ref{tab-translation}.

\par
The energy spectrum of b and c flavoured
hadrons affects both the b tagging and the hadronisation.
For b quark events the parameter of the Peterson et al.
fragmentation function \cite{bib-Peterson-fragmentation} used in the
JETSET generator \cite{bib-JETSET}, $\epsilon_{\rm b}$, was varied from
its default value of 0.0038 by $\pm 0.0010$, the range given in
\cite{bib-OPALtune,LEPEWWG-n_ch,LEPEWWG-tau}. Additionally two
different fragmentation functions were used, Kartvelishvili et al.
\cite{Kartvelishvili} and Collins and Spiller
\cite{Collins_Spiller}. The largest deviation between the three
variations and the standard result was taken as the uncertainty due to
the modelling of the b fragmentation. For c quark events the
parameter of the Peterson et al. fragmentation function,
$\epsilon_{\rm c}$,
was varied from its default value of 0.031 by $\pm$ 0.010
\cite{bib-OPALtune,LEPEWWG-n_ch,LEPEWWG-tau}.
\par
As shown in Figure \ref{fig-decaylength-significance}, the fraction of
b candidates in the data is slightly higher than in the Monte Carlo
prediction. To estimate the effect of this discrepancy on the
determination of the b quark mass, the tagging efficiencies
$\epsilon$, $\epsilon_3(\yc)$ were varied by a common constant
factor. The $1.5\%$ excess of b candidates in 
data compared to Monte Carlo leads to a negligible difference in the
mean value for $\mbMS$, as expected from the small uncertainty
associated with the variation of $R_{\Pb}$. No uncertainty was
assigned.

%%%% systematic errors in the double ratio
\begin{sidewaystable}
\centering
\begin{tabular}{|l||r|r|r|r|r|r|r|r|}
\cline{2-9}
\multicolumn{1}{c|}{ }
  & JADE
  & DURHAM
  & JADE E0
  & JADE P
  & JADE P0
  & JADE E
  & GENEVA
  & CAMBRI.
  \\
\cline{2-9}
\multicolumn{1}{c|}{ }
&
\multicolumn{8}{c|}{$B_3$}
\\
\cline{2-9}
\hline
result
  & $  0.9752 $
  & $  0.9532 $
  & $  1.0657 $
  & $  0.9938 $
  & $  1.0140 $
  & $  1.0649 $
  & $  0.9817 $
  & $  0.9717 $ \\
\hline
statistics
  & $ \pm 0.0048 $
  & $ \pm 0.0056 $
  & $ \pm 0.0035 $
  & $ \pm 0.0040 $
  & $ \pm 0.0035 $
  & $ \pm 0.0032 $
  & $ \pm 0.0094 $
  & $ \pm 0.0062 $ \\
\hline\hline
detector simulation
& $ \pm 0.0047 $
& $ \pm 0.0143 $
& $ \pm 0.0034 $
& $ \pm 0.0076 $
& $ \pm 0.0024 $
& $ \pm 0.0087 $
& $ \pm 0.0218 $
& $ \pm 0.0161 $
\\
 \hline
$|\cos(\theta_{\rm Thrust})| \le 0.75 ^{+0.05}_{-0.05}$
& $ \pm 0.0008 $
& $ \pm 0.0011 $
& $ \pm 0.0025 $
& $ \pm 0.0009 $
& $ \pm 0.0013 $
& $ \pm 0.0024 $
& $ \pm 0.0032 $
& $ \pm 0.0024 $
\\
\hline
$L/\sigma_L \ge 8 $
& $ \pm 0.0061 $
& $ \pm 0.0041 $
& $ \pm 0.0038 $
& $ \pm 0.0050 $
& $ \pm 0.0062 $
& $ \pm 0.0089 $
& $ \pm 0.0077 $
& $ \pm 0.0055 $
\\
 \hline
$ n_{\rm charged}^{\rm decay} = 4.955 \pm 0.062 $
& $ \pm 0.0002 $
& $ \pm 0.0010 $
& $ \pm 0.0004 $
& $ \pm 0.0005 $
& $ \pm 0.0003 $
& $ \pm 0.0006 $
& $ \pm 0.0004 $
& $ \pm 0.0006 $
\\
 \hline
$\tau_{\Pb}=(1.564 \pm 0.014)\;\mathrm{ps}$
& $ \pm 0.0000 $
& $ \pm 0.0005 $
& $ \pm 0.0003 $
& $ \pm 0.0003 $
& $ \pm 0.0002 $
& $ \pm 0.0003 $
& $ \pm 0.0006 $
& $ \pm 0.0005 $
\\
 \hline
$R_{\Pb}=(0.2175 ^{+0.013}_{-0.013})$
& $ \pm 0.0015 $
& $ \pm 0.0007 $
& $ \pm 0.0017 $
& $ \pm 0.0013 $
& $ \pm 0.0018 $
& $ \pm 0.0015 $
& $ \pm 0.0004 $
& $ \pm 0.0006 $
\\
\hline
b quark fragmentation
& $ \pm 0.0088 $
& $ \pm 0.0059 $
& $ \pm 0.0151 $
& $ \pm 0.0114 $
& $ \pm 0.0099 $
& $ \pm 0.0065 $
& $ \pm 0.0102 $
& $ \pm 0.0075 $
\\
 \hline
c quark fragmentation
& $ \pm 0.0007 $
& $ \pm 0.0001 $
& $ \pm 0.0008 $
& $ \pm 0.0003 $
& $ \pm 0.0003 $
& $ \pm 0.0001 $
& $ \mp 0.0015 $
& $ \mp 0.0005 $
\\
\hline
$b=(0.52 ^{+0.04}_{-0.04}) {\mathrm{GeV^2}}$
& $ \pm 0.0043 $
& $ \pm 0.0013 $
& $ \pm 0.0013 $
& $ \pm 0.0029 $
& $ \pm 0.0037 $
& $ \pm 0.0051 $
& $ \pm 0.0014 $
& $ \pm 0.0027 $
\\
\hline
$\sigma_q=(0.40 ^{+0.03}_{-0.03})$~GeV
& $ \pm 0.0017 $
& $ \pm 0.0012 $
& $ \pm 0.0009 $
& $ \pm 0.0012 $
& $ \pm 0.0011 $
& $ \pm 0.0018 $
& $ \mp 0.0012 $
& $ \pm 0.0009 $
\\
\hline
$Q_0=(1.90 ^{+0.50}_{-0.50})$~GeV
& $ \pm 0.0023 $
& $ \pm 0.0036 $
& $ \mp 0.0010 $
& $ \mp 0.0019 $
& $ \mp 0.0021 $
& $ \mp 0.0010 $
& $ \pm 0.0026 $
& $ \pm 0.0026 $
\\
\hline
$m_{\mathrm{b,JETSET}}=(5.0 \pm 0.5)$~GeV
& $ \pm 0.0002 $
& $ \pm 0.0000 $
& $ \pm 0.0011 $
& $ \mp 0.0001 $
& $ \pm 0.0011 $
& $ \mp 0.0014 $
& $ \pm 0.0003 $
& $ \mp 0.0004 $
\\
\hline
\hline
total systematic uncertainty
& $ \pm 0.0129 $
& $ \pm 0.0166 $
& $ \pm 0.0164 $
& $ \pm 0.0152 $
& $ \pm 0.0129 $
& $ \pm 0.0154 $
& $ \pm 0.0257 $
& $ \pm 0.0192 $
 \\
\hline
\end{tabular}
\
\caption{\label{tab-systematics}
          The systematic uncertainties in the value of the double ratio $B_3$.
          The correlations between the jet finders are taken into
          account.
          }

\end{sidewaystable}

%%%% systematic errors I
\begin{sidewaystable}
\centering
\begin{tabular}{|l||r|r|r|r|r|r|r|}
\cline{2-8}
\multicolumn{1}{c|}{ }
  & JADE
  & DURHAM
  & JADE E0
  & JADE P
  & JADE P0
  & JADE E
  & GENEVA
  \\
\cline{2-8}
\multicolumn{1}{c|}{ }
 & \multicolumn{7}{c|}{$\mbMS$~[GeV]}
\\
\cline{2-8}
\hline
result
  & $ 2.51 $
  & $ 3.20 $
  & $ 2.91 $
  & $ 1.73 $
  & $ 2.58 $
  & $ 2.47 $
  & $ 3.84 $
 \\
\hline
statistics
& $ \pm 0.42 $ 
& $ \pm 0.24 $ 
& $ \mp 0.10 $ 
& $ \mp 0.34 $ 
& $ \mp 0.12 $ 
& $ \mp 0.07 $ 
& $ \pm 0.55 $
 \\
 \hline\hline
detector simulation
  & $ \pm 0.35 $
  & $ \pm 0.55 $
  & $ \mp 0.10 $
  & $ \mp 0.72 $
  & $ \mp 0.08 $
  & $ \mp 0.19 $
  & $ \pm 1.04 $
 \\
 \hline
$|\cos(\theta_{\mathrm{Thrust}})| \le 0.75 ^{+0.05}_{-0.05}$
& $ \pm 0.06 $ 
& $ \pm 0.04 $ 
& $ \mp 0.07 $ 
& $ \mp 0.07 $ 
& $ \mp 0.04 $ 
& $ \mp 0.05 $ 
& $ \pm 0.17 $ 
 \\
 \hline
$L/\sigma_L \ge 8 $
  & $ \pm 0.54 $
  & $ \pm 0.18 $
  & $ \mp 0.11 $
  & $ \mp 0.35 $
  & $ \mp 0.20 $
  & $ \mp 0.18 $
  & $ \pm 0.44 $
 \\
 \hline
$ n_{\rm charged}^{\rm decay} = 4.955 \pm 0.062 $
  & $ \pm 0.01 $
  & $ \pm 0.04 $
  & $ \mp 0.01 $
  & $ \mp 0.04 $
  & $ \mp 0.01 $
  & $ \mp 0.01 $
  & $ \pm 0.02 $
 \\
 \hline
$\tau_{\Pb}=(1.564 \pm 0.014)\;\mathrm{ps}$
  & $ \pm 0.00 $
  & $ \pm 0.02 $
  & $ \mp 0.01 $
  & $ \mp 0.02 $
  & $ \mp 0.01 $
  & $ \mp 0.00 $
  & $ \pm 0.03 $
 \\
 \hline
$R_{\Pb}=(0.2175 ^{+0.013}_{-0.013})$
& $ \pm 0.11 $ 
& $ \pm 0.03 $ 
& $ \mp 0.05 $ 
& $ \mp 0.10 $ 
& $ \mp 0.06 $ 
& $ \mp 0.03 $ 
& $ \pm 0.02 $ 
 \\
 \hline
b quark fragmentation
  & $ \pm 0.62 $
  & $ \pm 0.23 $
  & $ \pm 0.46 $
  & $ \pm 1.73 $
  & $ \pm 0.36 $
  & $ \pm 0.14 $
  & $ \pm 0.52 $
 \\
 \hline
c quark fragmentation
& $ \pm 0.05 $ 
& $ \mp 0.00 $ 
& $ \mp 0.02 $ 
& $ \mp 0.02 $ 
& $ \mp 0.01 $ 
& $ \mp 0.00 $ 
& $ \pm 0.08 $ 
 \\
 \hline
$b=(0.52 ^{+0.04}_{-0.04}) {\mathrm{GeV^2}}$
& $ \pm 0.32 $ 
& $ \pm 0.05 $ 
& $ \mp 0.04 $ 
& $ \mp 0.24 $ 
& $ \mp 0.13 $ 
& $ \mp 0.11 $ 
& $ \pm 0.07 $ 
 \\
 \hline
$\sigma_{\rm q}=(0.40 ^{+0.03}_{-0.03})$~GeV
& $ \pm 0.14 $ 
& $ \pm 0.05 $ 
& $ \mp 0.03 $ 
& $ \mp 0.09 $ 
& $ \mp 0.04 $ 
& $ \mp 0.04 $ 
& $ \mp 0.06 $ 
 \\
 \hline
$Q_0=(1.90 ^{+0.50}_{-0.50})$~GeV
& $ \pm 0.18 $ 
& $ \pm 0.15 $ 
& $ \mp 0.03 $ 
& $ \mp 0.15 $ 
& $ \mp 0.07 $ 
& $ \mp 0.02 $ 
& $ \pm 0.14 $ 
 \\
 \hline
$m_{\mathrm{b,JETSET}}=(5.0 \pm 0.5)$~GeV
  & $ \pm 0.01 $
  & $ \pm 0.00 $
  & $ \pm 0.03 $
  & $ \mp 0.01 $
  & $ \pm 0.04 $
  & $ \mp 0.03 $
  & $ \pm 0.01 $
 \\
\hline
total systematic uncertainty
& $ \pm 0.99 $ 
& $ \pm 0.64 $
& $ \pm 0.50 $ 
& $ \pm 1.93 $ 
& $ \pm 0.45 $ 
& $ \pm 0.32 $ 
& $ \pm 1.27 $
 \\
\hline\hline
renormalisation scale
  & $ \pm 0.15 $
  & $ \pm 0.18 $
  & $ \pm 0.26 $
  & $ \mp 0.08 $
  & $ \pm 0.08 $
  & $ \pm 0.25 $
  & $ \pm 0.11 $
 \\
 \hline
$ \Delta a_0^i, \Delta a_2^i $
  & $ \pm 0.05 $
  & $ \pm 0.03 $
  & $ \pm 0.01 $
  & $ \pm 0.07 $
  & $ \pm 0.03 $
  & $ \pm 0.02 $
  & $ \pm 0.05 $
 \\
 \hline
$ \as=0.1184\pm0.0031$
  & $ \pm 0.02 $
  & $ \pm 0.01 $
  & $ \pm 0.03 $
  & $ \pm 0.03 $
  & $ \pm 0.01 $
  & $ \pm 0.03 $
  & $ \pm 0.01 $
 \\
\hline
total theoretical uncertainty
  & $ \pm 0.16 $
  & $ \pm 0.18 $
  & $ \pm 0.26 $
  & $ \pm 0.11 $
  & $ \pm 0.09 $
  & $ \pm 0.25 $
  & $ \pm 0.12 $
 \\
 \hline\hline
total uncertainty
& $ \pm 1.09 $ 
& $ \pm 1.20 $ 
& $ \pm 0.56 $ 
& $ \pm 1.86 $ 
& $ \pm 0.52 $ 
& $ \pm 0.41 $ 
& $ \pm 1.50 $ 
 \\
\hline
\end{tabular}
\caption{\label{tab-translation}
          The uncertainties in the value of the b quark mass at the \PZz\
          scale. The relative sign between uncertainties for different
          jet finders indicate the sign of their correlation coefficient.}
\end{sidewaystable}

\subsection{Hadronisation uncertainties}
\label{had_corr}
To assess the systematic uncertainties related to the hadronisation
process, 
HERWIG \cite{bib-HERWIG}
was tried as an alternative hadronisation model. It was found that for
this generator physics involving b quarks is not well described. Among
other problems the scaled mean energy of weakly decaying b mesons was
too low by several standard deviations. Also the number of tracks of
charged particles found per event was not modelled
correctly. Therefore no uncertainty was assigned.

To nevertheless assess the uncertainties related to hadronisation, 
we altered the main parameters of the JETSET model and
recalculated the hadronisation correction and the resulting value of
$\mb$. Beyond the variations affecting the b and c quark fragmentation
mentioned in the previous section, we altered the value of the $b$
parameter, which affects the hardness of the fragmentation function
for d,u and s quarks, the width $\sigma_{\rm q}$ of the transverse momentum
distribution, and the parameter $Q_0$ which serves as the cut-off for
the parton shower. These parameters were varied within their
uncertainties quoted in \cite{bib-OPALtune}. Furthermore, the b
quark mass inside JETSET was varied by up to $\pm 0.5$ GeV in 0.1 GeV
steps.
Since a different quark mass significantly affects the details of the
hadron generation, 
e.g. emission of soft gluons, which are sensitive to the dead cone
effect or the formation of b flavoured hadrons, 
the \Pb\ quark mass was kept fixed
at the JETSET default value of $5$~GeV throughout the
hadronisation process. The variation of the mass value was applied only
in the calculation of the first gluon radiation probability from the
b quark which employs the first order matrix element
\cite{bib-Ioffe-PL78B-277}. All these variations are listed in
Tables \ref{tab-systematics} and \ref{tab-translation}.

\subsection{Theoretical uncertainties}
Three contributions to the theoretical uncertainty were
considered. The coefficients in Eq. (\ref{b3_2_mb}) have
uncertainties because of finite statistics used in the Monte Carlo
integration program. Therefore we altered the coefficients by the
uncertainties listed in Table \ref{tab-coefficients} and reperformed
the analysis. This accounts for the uncertainty of the calculation and
the parametrisation of the mass dependence of the double ratio $B_3$.
 Second, the value of $\as$ was varied from its
world average of 0.1184 by its uncertainty of 0.0031
\cite{bib-Bethke-0004021}.
Third, the renormalisation scale factor $x_{\mu}$ in
Eq. (\ref{variation_mu}) was varied by factors of $\frac{1}{2}$ and
2. This last variation estimates the impact of neglected higher order
terms in the perturbation series. The results of these systematic
variations are given in Tables \ref{tab-systematics} and
\ref{tab-translation}.
\par
\begin{table}
\begin{center}
\begin{tabular}{|c|c|c|c|c|c|c|c||l|}
\cline{1-8}
  JADE
  & DUR.
  & J. E0
  & J. P
  & J. P0
  & J. E
  & GEN.
  & CAM.
  & \multicolumn{1}{c}{ } \\
\cline{1-8}
\hline
\hline
 1
 & 0.48
 & 0.88
 & 0.82
 & 0.79
 & 0.64
 & 0.21
 & 0.38
 &  JADE
\\ \hline
 \multicolumn{1}{c|}{}
 & 1
 & 0.45
 & 0.56
 & 0.47
 & 0.51
 & 0.51
 & 0.83
 & DURHAM
\\ \cline{2-9}
 \multicolumn{2}{c|}{}
 & 1
 & 0.79
 & 0.78
 & 0.60
 & 0.21
 & 0.38
 & JADE E0
\\ \cline{3-9}
 \multicolumn{3}{c|}{}
 & 1
 & 0.74
 & 0.69
 & 0.38
 & 0.51
 & JADE P
\\ \cline{4-9}
 \multicolumn{4}{c|}{}
 & 1
 & 0.60
 & 0.22
 & 0.38
 & JADE P0
\\ \cline{5-9}
 \multicolumn{5}{c|}{}
 & 1
 & 0.33
 & 0.42
 & JADE E
\\ \cline{6-9}
 \multicolumn{6}{c|}{}
 & 1
 & 0.63
 & GENEVA
\\ \cline{7-9}
 \multicolumn{7}{c|}{}
 & 1
 & CAMBRIDGE
\\ \cline{8-9}
\end{tabular}
\end{center}
\caption{\label{tab-correlation}
  Statistical correlations of the double ratio $B_3$ between the eight
  jet finders. The coefficients are different from the ones quoted in
  \cite{bib-SLAC-PUB-7915} where 3 and more jet events
  are used to determine the b quark mass. }
\end{table}

\section{Combined result}
\label{sec-combination}
\begin{figure}
\begin{center}
  \mbox{\includegraphics[width=1.0\linewidth]{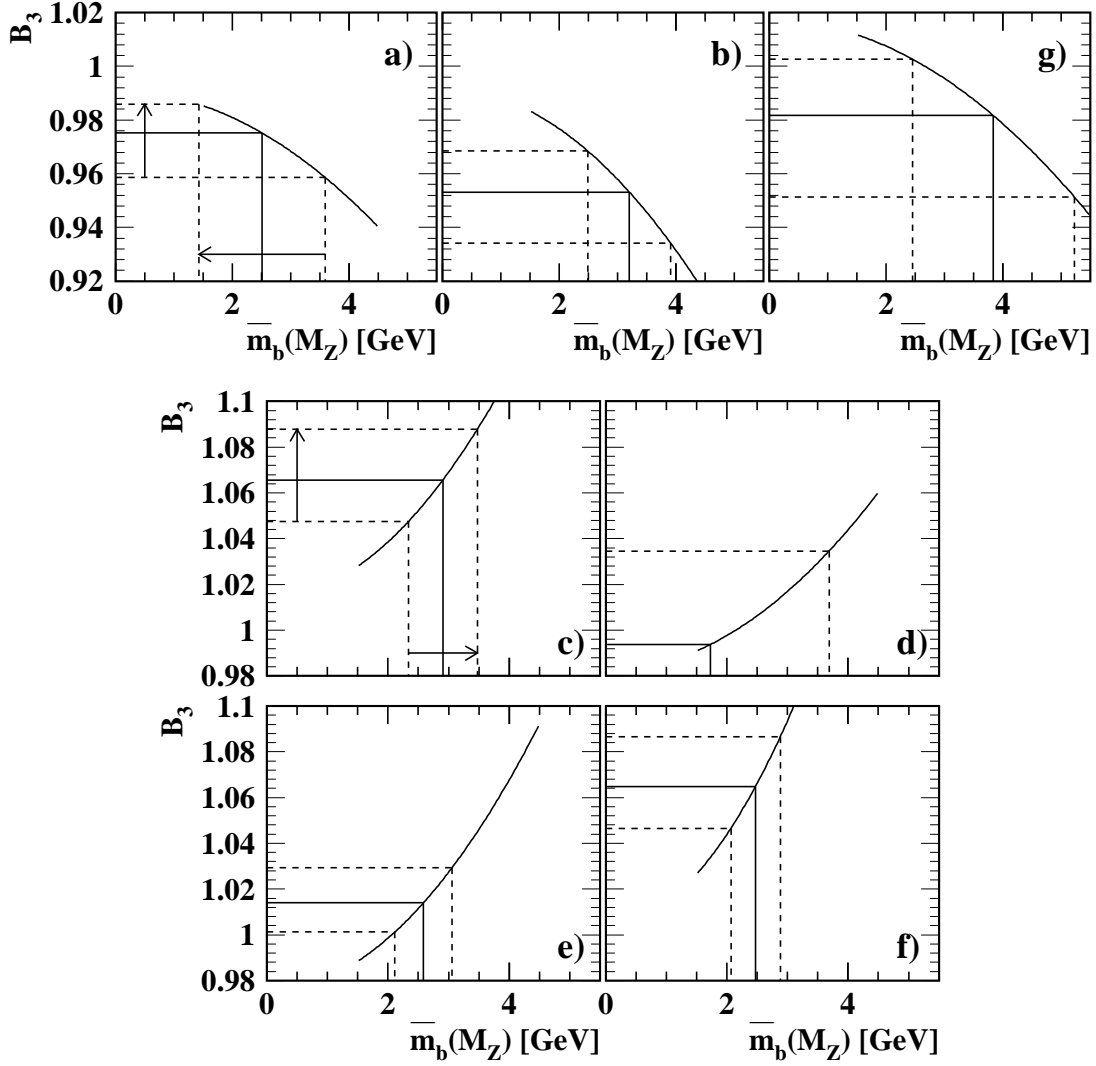}}
\end{center}
\caption{\label{fig6}
  For each jet finder the fit to the theoretical prediction,
  Eq. (\ref{b3_2_mb}), is shown together with each measured double
  ratio $B_3$ displayed on the y-axis.
  The b quark mass can be read from the x-axis. The band represents
  the total uncertainty for the double ratio and b quark mass.
  In a) JADE, b) DURHAM, c) JADE E0, d) JADE P, e) JADE P0, f) JADE E
  and in g) GENEVA schemes are shown. No lower bound on the b quark mass
  for the JADE P scheme is shown as this bound reaches the physical
  limit of a vanishing b quark mass.
%%% WITH ARROWS:
  The arrows in a) and c) indicate, how two positively correlated
  systematic uncertainties on $B_3$ turn into negatively correlated
  ones in $\mbMS$.
 }
\end{figure}
\begin{figure}
\begin{center}
  \mbox{\includegraphics[width=1.0\linewidth]{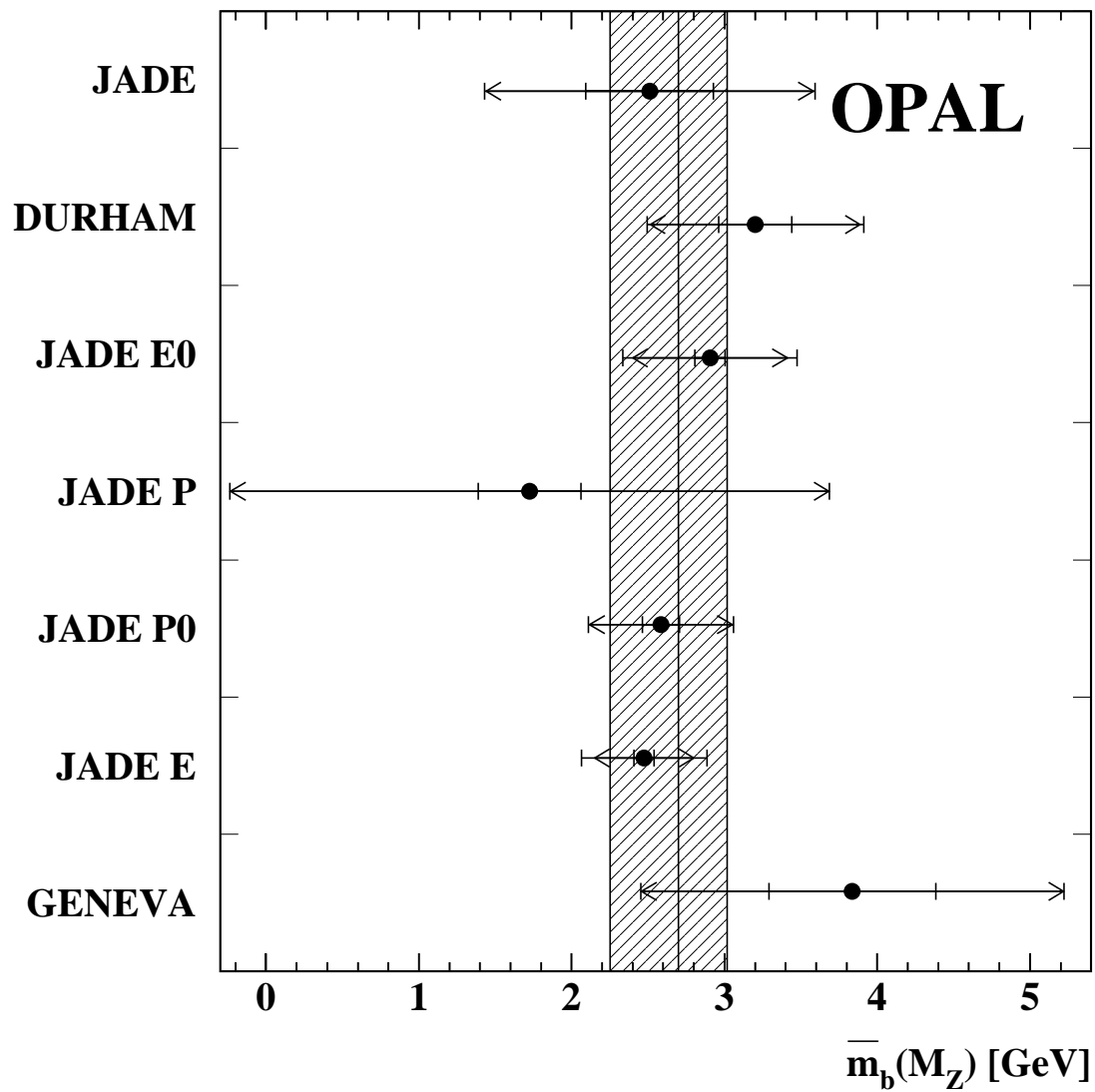}}
\end{center}
\caption{\label{result}
  The result of the combination of the seven individual measurements
  is shown by the vertical line, the hatched band displays its total
  uncertainty. Also shown are the mass values with their 
  uncertainties for each individual jet finder. The inner error bars
  are the statistical uncertainty. The arrows depict the statistical
  and systematic uncertainties added in quadrature, and the outer
  error bars depict the total uncertainty. }
\end{figure}

As can be seen in Table \ref{tab-translation} and in Figures
\ref{fig6} and \ref{result}, the individual results of all jet
finders agree
well within their total uncertainty. Hence we now consider a
combination of the seven determinations.
To account for correlations between the seven jet finders, the mean
was determined using a correlation matrix. This matrix was constructed
by dividing both data and Monte Carlo into 200 independent
subsamples\footnote{The inclusive sample contains about one million
events and allows for such a fine subdivision},
calculating for each subsample and each jet finder the double ratio
$B_3$ and determining the correlation between the jet finders. This
gave the statistical correlation matrix in Table \ref{tab-correlation}.
It differs from that given in \cite{bib-SLAC-PUB-7915} as here only
3-jet events are analysed, in contrast to 3 and more jet events in the
latter analysis.
Using the covariance matrix obtained from the correlation matrix, the
mean mass value $\overline{m}_{\mathrm{b}}$ of the seven mass
values was calculated by minimising
\begin{equation}
\chi^2 = \sum_{i,j}
[B_3^{(i)}-B_3^{(i),theo} (\overline{m}_{\mathrm{b}})]
\cdot [{\rm cov}^{-1}]_{ij} \cdot
[B_3^{(j)}-B_3^{(j),theo} (\overline{m}_{\mathrm{b}})]
\label{mini_chi}
\end{equation}
with $i=\mathrm{JADE}, \mathrm{E}, \mathrm{E0}, \mathrm{P}, \mathrm{P0}, 
\mathrm{DURHAM}, \mathrm{GENEVA}$ and ${\rm cov}$ the covariance
matrix of all jet finders $i$. This yielded the result
\begin{math}
\chiSquareN \; ,
\end{math}
with a statistical uncertainty of $\pm 0.03 {\mathrm{~GeV}}$ and
a $\chi^2/\mathrm{d.o.f.}$ of $100/6$.
A large $\chi^2/\mathrm{d.o.f.}$ can be expected since at this stage
only statistical uncertainties are accounted for. This has been seen
also in earlier studies \cite{bib-SLAC-PUB-7915}.
\par
To determine the effect of each of the
systematic variations considered in Section \ref{sec-systematics} on
the mean value, the diagonal elements of the covariance matrix for the
chosen variation were added to the statistical covariance
matrix\footnote{As in \cite{pr075} the covariance matrix for a
  systematic variation of an observable $O$ was constructed by
  assigning the product of two uncertainties $\delta O_i \cdot \delta
  O_j$ from two single measurements $i$ and $j$ to the matrix.}. With
this covariance matrix a new mean was calculated with the mass values
derived by this particular variation under consideration as input
values. Its deviation from the standard result was considered as the
systematic uncertainty due to this variation. 
The total uncertainty was calculated in the same way
as for each individual jet finder. Using the same covariance matrix of
statistical correlations for all systematic variations assumes that
the matrix changes only slightly for different variations.
The theoretical uncertainty was calculated using the weighted mean
method.
This procedure yields a combined result of
\begin{equation}
 \ExtendedWeightedMean\;.
\end{equation}
\par
The uncertainty labelled ''syst.'' includes the detector and the
hadronisation terms in Tables \ref{tab-systematics} and
\ref{tab-translation}.
The stability of this result was tested by calculating the mean and
uncertainties using any combination of 2,3,4,5 or 6 out of the 7 jet
finders and no large deviations from the standard analysis was found.
\par
To estimate the effect of a non-vanishing c quark mass, 
the parton level of the light quark sample was modified by replacing c
quark events by u quark events and
repeating the analysis. The negligible shift of the mean compared to
the standard result was $+0.02~\mathrm{GeV}$, which was added in
quadrature to the systematic uncertainty.
A weighted mean was also calculated
and gave consistent results.
\par
Figure \ref{result} shows the b quark mass for the seven individual
jet finders together with the mean value of the combination and its
total uncertainty.
\section{Summary}
\label{sec-conclusions}
\begin{figure}
\begin{center}
  \mbox{\includegraphics[width=1.0\linewidth]{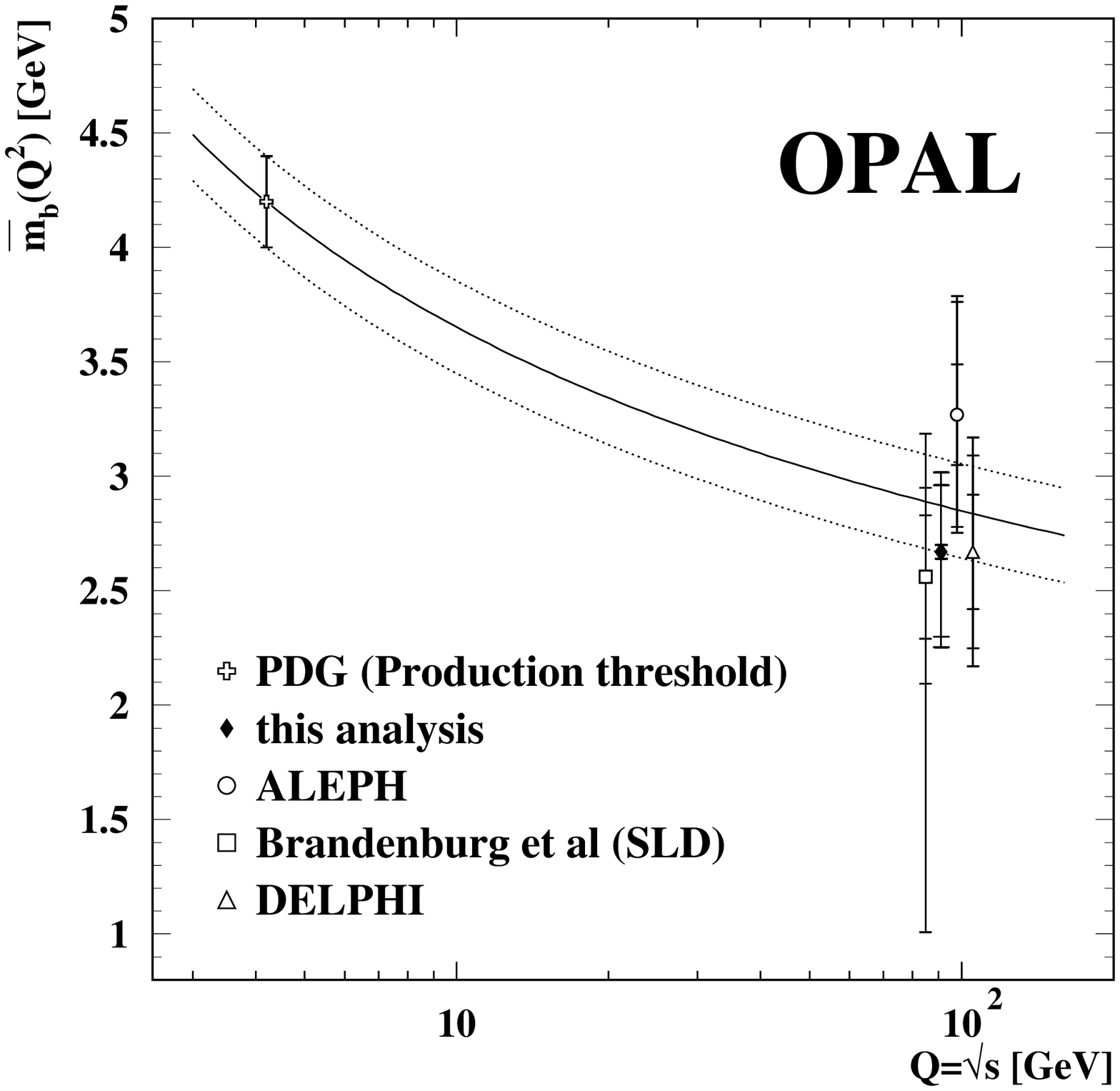}}
\end{center}
\caption{\label{fig-running-mb}
  Results for $\mbMS(\mZ)$ from this analysis and at the
  scale \mbMS\ from the compilation in 
  \protect\cite{RPP2000}, together with other determinations at the
  $\PZz$ mass scale from DELPHI
  \cite{bib-mb-at-mZ}, 
  ALEPH \cite{ALEPH_mb} and
  Brandenburg et al. \cite{bib-SLAC-PUB-7915}.
  The solid curve shows the theory expectation
  for the running of the b quark mass, Eq.(\ref{eqn-running-mb}),
  using $\as(\mZ^2) = 0.1184$ 
  \cite{bib-Bethke-0004021}
  and $\mbMS^{(\mathrm{PDG})}(\mbMS) = 4.2 {\mathrm{~GeV}}$
  \cite{RPP2000}. The dotted lines display the total
  uncertainty on the b quark mass run from production threshold to the
  $\PZz$ pole. This total uncertainty includes the
  uncertainty on the mass itself, $\mbMS^{(\mathrm{PDG})}(\mbMS) =
  (4.2 \pm 0.2) {\mathrm{~GeV}}$ and the uncertainty on
  $\as(\mZ^2) = 0.1184 \pm 0.0031$.
  The error bars of the data points show the statistical, the
  systematic and the theoretical uncertainties added in quadrature.
  For displaying purposes the four measurements at the Z pole have
  been separated.
  }
\end{figure}

The b quark mass was determined at the \PZz\ mass scale by
comparing the $3$-jet rates in b and light quark events using seven
different jet finders.
A deviation of the $3$-jet rates in tagged b events compared with light
quark events was observed. This deviation was used to derive the
b quark mass by comparing to the theoretical prediction.
By minimising the $\chi^2$ of the seven correlated determinations a
single result for the b quark mass was determined to be
\begin{equation}
 \ExtendedWeightedMean\;.
\end{equation}
Our final result is shown in Figure \ref{fig-running-mb}. Also shown
in this figure
is the average of the b quark mass at the scale of the b quark mass
itself, \mbox{$\mbMS^{(\mathrm{PDG})}(\mbMS)=(4.2\pm 0.2)$~GeV.} as
compiled in \cite{RPP2000}.
This average has been derived from measurements at the \bbbar\ 
production threshold and from \Pb\ hadron masses. Also shown are
other measurements of $\mbMS(\mZ)$ by 
DELPHI \cite{bib-mb-at-mZ}, ALEPH \cite{ALEPH_mb} and
Brandenburg et al. using SLD data \cite{bib-SLAC-PUB-7915}. These
determinations at the $\PZz$ mass scale yielded mass values in the
range of $2.67$~GeV to $3.27$~GeV. All results are in good agreement
with each other. The total uncertainty on the b quark mass in this
analysis is smaller than for previous measurements at the \PZz\ mass
scale, see \mbox{Figure \ref{fig-running-mb}.}
The solid curve in Figure \ref{fig-running-mb} shows the QCD
expectation of a scale dependent \Pb\ quark mass in the
\MS~renormalisation scheme, using the world average value
of $\as(\mZ^2)=0.1184 \pm 0.0031$ \cite{bib-Bethke-0004021}.
Evolving our result of $\mbMS(\mZ)$ down to the b quark mass scale
itself gives 
\begin{displaymath}
  \evolvedToThreshold\;,
\end{displaymath}
which is in agreement with $(4.2 \pm 0.2) {\mathrm{~GeV}}$ \cite{RPP2000}.

We have also compared our result for the b quark mass at the $\PZz$ mass
scale with the combined value at production threshold, yielding
\begin{equation}
  \mbMS^{(\mathrm{PDG})}(\mbMS) - \mbMS(\mZ) = (1.53 \pm 0.39) 
                                  {\mathrm{~GeV}}\;,
\end{equation}
which is different from zero by 3.9 standard deviations, confirming
the running of the b quark mass in the \MS~ renormalisation scheme
as predicted by QCD. 

%%%%%%%%%% Start standard additions %%%%%%%%%%
\appendix
\par
Acknowledgements:
\par
We thank A.Brandenburg for valuable discussion.
We particularly wish to thank the SL Division for the efficient operation
of the LEP accelerator at all energies
 and for their continuing close cooperation with
our experimental group.  We thank our colleagues from CEA, DAPNIA/SPP,
CE-Saclay for their efforts over the years on the time-of-flight and trigger
systems which we continue to use.  In addition to the support staff at our own
institutions we are pleased to acknowledge the  \\
Department of Energy, USA, \\
National Science Foundation, USA, \\
Particle Physics and Astronomy Research Council, UK, \\
Natural Sciences and Engineering Research Council, Canada, \\
Israel Science Foundation, administered by the Israel
Academy of Science and Humanities, \\
Minerva Gesellschaft, \\
Benoziyo Center for High Energy Physics,\\
Japanese Ministry of Education, Science and Culture (the
Monbusho) and a grant under the Monbusho International
Science Research Program,\\
Japanese Society for the Promotion of Science (JSPS),\\
German Israeli Bi-national Science Foundation (GIF), \\
Bundesministerium f\"ur Bildung und Forschung, Germany, \\
National Research Council of Canada, \\
Research Corporation, USA,\\
Hungarian Foundation for Scientific Research, OTKA T-029328, 
T023793 and OTKA F-023259.\\
%\end{document}
%%%%%%%%%%%%%%%%%%%%%%%%%%%%%%%%%%%%%%%%
%
%\newpage

\end{document}